\documentclass[aps,twocolumn,showpacs,showkeys]{revtex4}
\usepackage{graphicx}

\begin{document}

\title{Surface-Enhanced Raman Scattering (SERS) and Surface-Enhanced Fluorescence (SEF) in the context of modified spontaneous emission}

\author{E. C. Le Ru} \email{Eric.LeRu@vuw.ac.nz}

\author{P. G. Etchegoin} \email{Pablo.Etchegoin@vuw.ac.nz}

\affiliation{The MacDiarmid Institute for Advanced Materials and
Nanotechnology, School of Chemical and Physical Sciences, Victoria
University of Wellington, PO Box 600, Wellington, New Zealand}
\homepage[]{http://www.vuw.ac.nz/Raman}

\date{\today}

\begin{abstract}
Surface Enhanced Raman Scattering (SERS) and Surface-Enhanced
Fluorescence (SEF) are studied within the framework of modified
Spontaneous Emission (SE), and similarities and differences are
highlighted. This description sheds new light into several aspects
of the SERS electromagnetic enhancement. In addition, combined
with the optical reciprocity theorem it also provides a rigorous
justification of a generalized version of the widely used SERS
enhancement factor proportional to the fourth power of the field
($|E|^4$). We show, in addition, that this approach also applies
to the calculation of Surface-Enhanced Fluorescence cross-sections
thus presenting both phenomena SERS and SEF within a unified
framework.
\end{abstract}



\maketitle

\section{Introduction}

The emission properties of a dipole are affected by its
environment. This fact was first pointed out by Purcell
\cite{Purcell46} and has since then been demonstrated
experimentally in a wide range of experimental situations; for
example for excited Eu$^{3+}$ ions in front of a plane metallic
surface, as in the original observation by Drexhage
\cite{Drexhage68}, or for Rydberg atoms in an optical cavity
\cite{Goy83}. This effect can be particularly strong in close
proximity of metallic structures \cite{Chance78,Metiu84,Ford84}.
It is evident in the modification (quenching or enhancement) of
spontaneous emission (SE) rates \cite{Dulkeith02}. Although this
effect has been assumed for some time to lead to a strong
quenching of the fluorescence, instances of enhanced fluorescence
have been reported more recently for dyes \cite{Lakowicz05}, and
even for quantum dots \cite{Song05}, so-called Surface-Enhanced
Fluorescence (SEF) or Metal-Enhanced Fluorescence (MEF). Such
enhancements have been known for some time and are even more
spectacular in Surface Enhanced Raman Scattering (SERS)
\cite{Moskovits85}. SERS has regained interest lately following
the observation of single molecules \cite{Kneipp97,Nie97}, and
possible applications in biology/analytical chemistry
\cite{VoDinh99,Kneipp02} and nano-plasmonics \cite{LeRuCAP05}.
SERS enhancement factors as large as $10^{15}$ have been reported
and are believed to be mainly electromagnetic (EM) in origin,
although the detailed mechanism is still debated. It is widely
accepted that localized surface plasmon resonances give rise to
places with large local field enhancement factors
$|E_{\text{Loc}}/E_0|$; so-called hot-spots. In most cases the
SERS signal is assumed to be proportional to the power of 4 of
this factor \cite{Kneipp02,Jiang03,Xu04}: a power of two
accounting for the excitation part, and a power of two for the
emission. This assumption, although widely used in particular to
explain the enhancements required for single molecule SERS, has
never been rigourously justified, as we shall show in the
forthcoming discussion.

 Numerous theoretical studies have concentrated on the modification of SE rates inside
dielectric materials, perfect or absorbing, and close to
boundaries with dielectrics or metals. Optical cavities, and
similarly photonic crystals, can also strongly enhance, suppress,
or redirect the SE field of an atom \cite{Goy83}. Most available
results in the literature relate to SE rates in simple geometries
such as plane boundaries, spheres, or ellipsoids. Only recently,
methods have been proposed for SE rates in more complex
nano-structures \cite{Blanco04}. Scattering processes (elastic or
inelastic) by an atom are in many ways similar to spontaneous
emission. Excluding stimulated processes, the scattering of a
photon by an atom involves the `spontaneous' creation of a photon
in a mode where no photon existed before. One therefore expects a
modification of the scattering cross-section in a similar way as
SE rates are modified by the environment. The difference between
these two processes only lies in the initial state, to wit:
excited atom with no photon for SE, or atom in ground state with
one incident photon for the case of scattering. Fluorescence is
also in many respects similar to scattering. The crucial
difference is that scattering is instantaneous while fluorescence
is a stepwise process involving absorption, energy relaxation, and
spontaneous emission.

It is one of the purposes of this paper to highlight the
similarities, and discuss the differences between SE, SEF, and
SERS for atoms or molecules in close proximity of metal surfaces.
These aspects have been discussed in reviews, in the past
\cite{Chance78,Metiu84,Ford84} and more recently
\cite{Lakowicz05}, but usually focusing on one specific process.
The emphasis will be put, therefore, on the relationships between
these processes presenting them under one single framework in a
more modern context. In particular, the various methods for
calculating SE rates and cross sections, some of them common and
others more novel, are described in detail. The description is
kept as simple as possible to emphasize the physical aspects. For
this reason, we constrain our description to classical
electrodynamics with a local dielectric function. We also use
isotropic polarizabilities whenever necessary to avoid complicated
tensorial notations. The results are illustrated by simple
examples of emitters close to a silver sphere. The main advantage
is that it does not rely on any approximation for the solution of
the electromagnetic problem, since analytic expressions (not
discussed here) can be obtained from Mie theory \cite{Mie08}.
These choices do not affect the generality of the results and the
methods described here can be used for complex geometries provided
that a suitable method --usually numerical-- is available to solve
the electromagnetic problem.

Finally, particular emphasis will also be given to the EM
enhancements in SERS. We derive a general formula for the SERS
cross sections in the framework of modified SE rates. It
highlights the importance of non-radiative vs. radiative
enhancements and also emphasizes the strong similarities between
SERS enhancement and enhanced SE which, by the same token,
requires to reconsider the origin of the fluorescence quenching
under SERS conditions. In addition,
 the optical reciprocity theorem \cite{Landau} is used to
show that these results are consistent with a generalized version
of the commonly used $|E|^4$ factor in SERS, with certain
(generally overlooked) limitations.

The paper is organized as follows: in Sec. \ref{SecSE}, we discuss
modified SE rates and the various ways of calculating them. In
particular, we highlight the use of the optical reciprocity
theorem as a method for calculating SE rates, a technique that has
not received the attention it should deserve. The modification of
the quantum yield of a fluorophore is also discussed. In Sec.
\ref{SecSERS}, we apply these arguments to the determination of
SERS cross-section. We use the reciprocity theorem to derive a
generalized version of the common $|E|^4$ enhancement factor, and
show a simple example where this factor would lead to erroneous
results. Similar arguments are applied to obtain a simple
description of SEF. It is shown that fluorescence quenching or
enhancement are usually both possible, depending on the relative
contribution of two competing mechanisms, namely: absorption
cross-section enhancement and non-radiative quenching. Finally in
Sec. \ref{SecDisc}, we discuss the differences and similarities
between SE rates and SERS cross sections. Some common aspects of
SERS, such as fluorescence quenching or the presence of an
inelastic continuum, and other outstanding issues (such as
vibrational pumping) are revisited within this new approach.

\section{Modified Spontaneous Emission}
\label{SecSE} Spontaneous emission is an intrinsically quantum
effect, where a dipolar atomic transition couples to the vacuum
state of the electromagnetic field. This coupling can only occur
with the quantized electromagnetic field, where the vacuum state
has a zero-point energy and exhibit quantum fluctuations. However,
because of the difficulties in the quantization of the field in
complex structures, the modification of SE rates is most often
studied within a semi-classical model of an oscillating
electromagnetic dipole \cite{Chance78}. This leads to results in
agreement (in most cases) with experiments and with the full
quantum mechanical treatment when possible. Several complementary
techniques can also be used; the first of which considers the
self-reaction field and is closest to the quantum treatment (and
can be shown to be formally equivalent in most cases). The latter
enables the derivation of a decay rate but does not distinguish
between radiative and non-radiative processes. To avoid this
problem, one can use a second technique based on the Poynting
vector, which emphasizes energy conservation. We review here the
derivation and main results of these two methods. We then describe
a third approach, which has hardly been used in the past, but will
be important in connection with the SERS cross sections.

SI units are used throughout the paper. Results from these three
approaches are illustrated by considering a dipolar emitter at a
short distance $d=1$ or 2\,nm from the surface of a silver sphere.
Standard Mie theory \cite{Mie08,Bohren83,Chew87} was used to
derive these results, and the environment is assumed to be vaccuum
($\epsilon_r=1$). The (local) wavelength-dependent relative
dielectric function of silver was approximated by:
\begin{equation}
\epsilon(\lambda)= \epsilon_{\infty} - \frac{1}
{\lambda_{p}^2~\left(\frac{1}{\lambda^2}+\frac{i}{\Gamma\lambda}\right)},
\label{Ag}
\end{equation}
where $\epsilon_{\infty}=4$, $\lambda_{p}=141$ nm, $\Gamma=17000$
nm. These parameters provide the best Drude fit for the real
optical properties of Ag.

\subsection{Spontaneous Emission and self-reaction}
\label{SecSR}

Within the classical treatment, SE can be seen as an effect of the
self-reaction, i.e. the electromagnetic field
$\mathbf{E}_{\text{SR}}$ created at the dipole position by itself,
either directly or through its interaction with the environment
(reflected field). We consider a dipole
$\mathbf{d}=d\mathbf{e}_d$, oscillating at a frequency $\omega$
(an $e^{-i\omega t}$ dependence is assumed for the fields). The
self-reaction field can be written as
$\mathbf{E}_{\text{SR}}=\mathbf{G}(\omega)\mathbf{d}$, where
$\mathbf{G}(\omega)$ can be viewed as the {\it self-reaction
Green's tensor}, obtained by solving Maxwell's equation with the
appropriate boundary conditions.

The SE rate can then be calculated from \cite{Barnett92,Blanco04}
\begin{equation}
\Gamma=\frac{2}{\hbar}d^2
\text{Im}(\mathbf{e}_d\mathbf{G}(\omega)\mathbf{e}_d),
\label{EqnGammaModQu}
\end{equation}
which can be derived from {\it quantum} arguments
\cite{Barnett92}. The parameter $d$ should be taken to be the
dipole moment of the transition
$d_{\text{qu}}^2=\left|-e\left<b\right|\mathbf{r} \cdot
\mathbf{e}_d\left|a\right>\right|^2$ between excited
$(\left|b\right>)$ and ground state $(\left|a\right>)$. However,
$\Gamma$ can be evaluated using Eq. \ref{EqnGammaModQu} from the
{\it classical} calculation of the Green's tensor
\cite{Barnett92}. For example, for an atom in free space, it can
be shown \cite{CDG1} that
$\mathbf{G}_0(\omega)=G_0(\omega)\mathbf{1}$, with
\begin{equation}
\text{Im}(G_0(\omega))=\frac{\omega^3}{6\pi\epsilon_0 c^3},
 \label{EqnImG0}
\end{equation}
 which leads to the usual spontaneous emission lifetime:
\begin{equation}
\Gamma_0=\frac{\omega_0^3 d_{\text{qu}}^2} {3\pi\epsilon_0 \hbar
c^3}. \label{EqnGammaQu}
\end{equation}
The reason why the classical self-reaction Green's tensor
$\mathbf{G}(\omega)$ can be used to determine $\Gamma$ is that it
is intricately linked to the vacuum fluctuations through the
quantum fluctuation-dissipation theorem \cite{Barnett92}.

In the presence of boundaries, in particular close to metallic
surfaces, the self-reaction field and therefore the SE rate can be
strongly modified. One can then write
$\mathbf{G}=\mathbf{G}_0+\mathbf{G}_r$, where $\mathbf{G}_r$
corresponds to the self-reaction due to the field reflected by the
boundaries at the dipole position.  {\it This can be calculated by
solving Maxwell's equation for the dipole emission in the presence
of boundaries}. The SE rate enhancement factor is then derived
from Eq. (\ref{EqnGammaModQu}) and reads
\begin{equation}
M_{\text{Tot}}(\omega)=\frac{\Gamma}{\Gamma_0}=1+\frac{\text{Im}
(\mathbf{e}_d\mathbf{G}_r(\omega)\mathbf{e}_d)}{\text{Im}(G_0(\omega))},
\label{EqnMTot}
\end{equation}
where $\text{Im}(G_0(\omega))$ is given by Eq. (\ref{EqnImG0}).

It is important to stress that $M_{\text{Tot}}$ here corresponds
to the modification of the {\it total decay rate} of the emitter.
$M_{\text{Tot}}$ can also be rewritten as $P_{\text{Tot}}/P_0$,
where $P_0$ is the power emitted in free space, and
$P_{\text{Tot}}$ is the total power extracted from the dipole in
the presence of the metal. In most situations, especially with
metals, only a fraction, $P_{\text{Rad}}$, of this power is
radiated in the far-field, while the remainder, $P_{\text{NR}}$,
is absorbed by the environment. In most measurements, only the
{\it radiative enhancement} is measured: $P_{\text{Rad}}/P_0$.
However, the emitter does experience a total decay rate
enhancement given by $M_{\text{Tot}}$. This can in principle only
be observed directly in a time-resolved experiment
\cite{Dulkeith02}. Moreover, a large value of $M_{\text{Tot}}$
does not necessarily imply a large radiative enhancement, in
particular in the presence of absorbing media such as metals
\cite{Dulkeith02,Barnett92,Blanco04}. The main inconvenience of
the approach described above is therefore that it does not lead to
any information about the radiative properties.

\begin{figure}
\centering{
 \includegraphics[width=8cm]{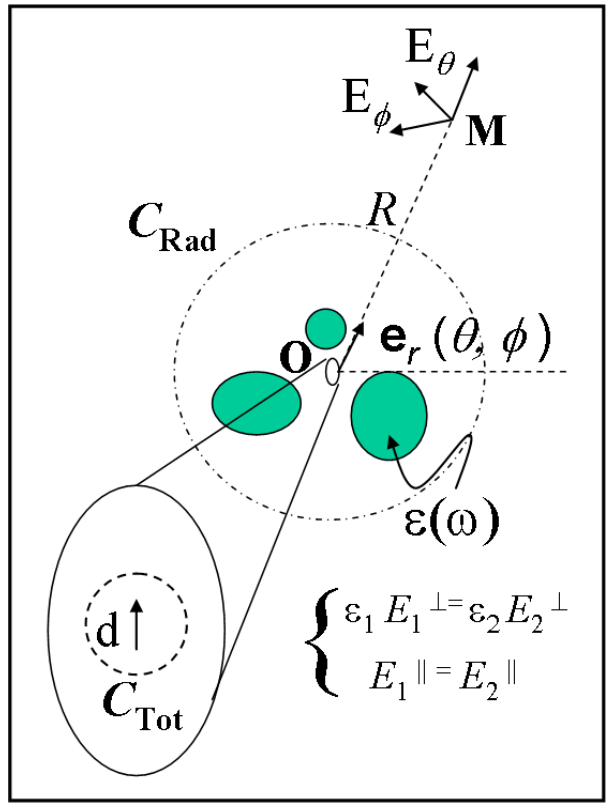}
} \caption{Schematic representation of the EM problem of an
emitter (dipole) in close proximity to a metallic environment. The
surfaces $C_{\text{Tot}}$ and $C_{\text{Rad}}$ are used to
evaluate $P_{\text{Tot}}$ and $P_{\text{Rad}}$ within the Poynting
vector approach. In the top-right corner we depict the radiation
field and its components, which will play a role for the
formulation of the optical reciprocity theorem. See the text for
further details.}
 \label{FigSchemaSE}
\end{figure}

\subsection{The Poynting vector approach}
\label{SecPoynting}

Another possible semi-classical approach is to directly calculate
the power $P_{\text{Tot}}$ extracted from a classical dipole
$d_{\text{cl}}$ and deduce
\begin{equation}
M_{\text{Tot}}(\omega)=\frac{P_{\text{Tot}}(\omega)}{P_0(\omega)},
 \label{EqnMTotPtot}
\end{equation}
 where $P_0=\omega^4 d_{\text{cl}}^2
/(12\pi \epsilon_0 c^3)$ is known from classical electrodynamics
\cite{Jackson}. $P_{\text{Tot}}$ can be estimated by solving
Maxwell's equations for the dipole in the presence of the
boundaries and then integrating the flux of the Poynting vector,
$\mathbf{S}$, through a closed surface outside the dipolar source
\cite{Chance78,Chew87,Blanco04}. Let us consider a dipolar emitter
(in vacuum) located close (but not inside) to one or several
finite size metallic objects. The situation is depicted
schematically in Fig. \ref{FigSchemaSE}. We can draw a surface
around the emitter, $C_{\text{Tot}}$, small enough to enclose no
absorbing media, and calculate the flux of the Poynting vector,
$P_{\text{Tot}}$, through this surface. Because of energy
conservation (no absorption inside this surface), $P_{\text{Tot}}$
must be equal to the power extracted from the dipole by the EM
field $P_{\text{Tot}}=-0.5\times \text{Re}(i\omega \mathbf{d}^{*}
\cdot \mathbf{E})$. From this expression and using the definition
of $\mathbf{G}$ and the value for $P_0$, one sees that Eq.
(\ref{EqnMTotPtot}) is exactly equivalent to Eq. (\ref{EqnMTot})
obtained from the self-reaction approach. The two approaches are
therefore fully consistent owing to energy conservation (when the
dipole is assumed to be located in a non-absorbing medium). Note
that the self-reaction approach, thanks to its direct link with
the quantum treatment, can be used to calculate the absolute decay
rate, $\Gamma$ or $\Gamma_0$, as in Eq. (\ref{EqnGammaQu}). The
Poynting vector approach, however, only yields the relative decay
rate (enhancement factor), $\Gamma/\Gamma_0$. This is not a
limitation here, since we are interested precisely in this decay
rate enhancement.

Besides, the Poynting vector approach also yields additional
information, namely: the non-radiative decay, the radiated power,
and its angular dependence (radiation pattern)
\cite{Chance78,Blanco04}. In the situation of Fig.
\ref{FigSchemaSE}, where the dipole environment is bounded, one
can enclose the emitter and all absorbing media in a large sphere,
$C_{\text{Rad}}$, and calculate the power outflow on this surface,
which corresponds to the power radiated by the source in the
far-field, $P_{\text{Rad}}$. Moreover, if this surface is
sufficiently far from any objects and sources, such that the lines
of the Poynting vector are approximately radial (radiation field),
then one can also obtain $dP_{\text{rad}}/d\Omega
(\theta,\phi)=r^2 \text{Re}(\mathbf{S}\cdot \mathbf{e}_r)$, i.e
the power radiated per unit solid angle, and deduce the angular
radiation pattern. Finally, from the solution of the EM field
inside the absorbing media, it is also possible to calculate the
total absorbed power, $P_{\text{NR}}$, which corresponds to
non-radiative losses. Alternatively, one can also use energy
conservation, which ensures that
$P_{\text{Tot}}=P_{\text{Rad}}+P_{\text{NR}}$, to obtain the
absorbed power from the other two magnitudes.

This technique therefore enables us to calculate the radiative
enhancement $M_{\text{Rad}}=P_{\text{Rad}}/P_0$, and the radiative
efficiency $\eta_{\text{Rad}}=P_{\text{Rad}}/P_{\text{Tot}}$.
These two quantities can also be defined and calculated as a
function of angle $(\theta,\phi)$. It is interesting to note that
a large radiative enhancement can coexist with a small radiative
efficiency, and vice-versa, depending on the weight of
non-radiative processes. What is measured in experiments is
usually either the {\it radiative} SE enhancement
$M_{\text{Rad}}$, or the {\it angle dependent radiative}
enhancement $M_{\text{Rad}}(\theta, \phi)$. The {\it total} SE
rate enhancement or decay rate $M_{\text{Tot}}$ (radiative+
non-radiative) can only be measured in time-resolved experiments.

\begin{figure}
\centering{
 \includegraphics[width=8cm]{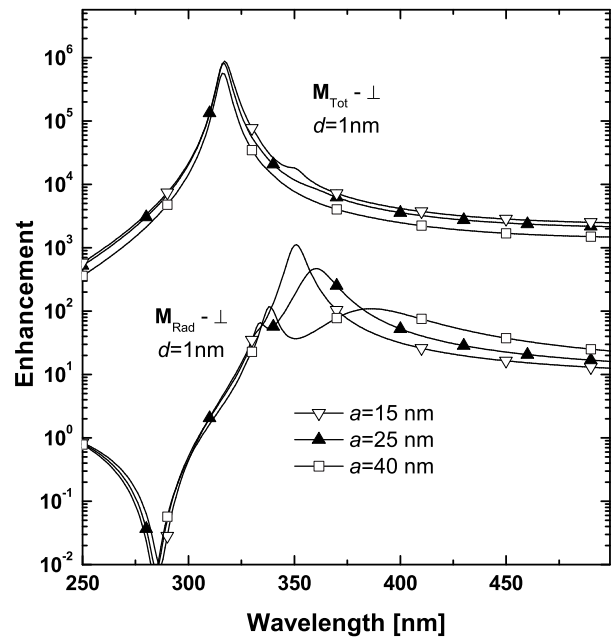}
} \caption{Total ($M_{\text{Tot}}$) and radiative
($M_{\text{Rad}}$) SE rate enhancements for a dipole perpendicular
(at a distance $d=1$~nm) to silver spheres of radii $a=$15, 25 and
40\,nm. This illustrates the influence of surface plasmon
resonances. The peak in $M_{\text{Rad}}$ red-shifts with
increasing $a$ and corresponds to the surface plasmon resonance of
the sphere, which is geometry-dependent. Note also that
enhancements are larger at resonance for the smallest spheres. The
peak in $M_{\text{Tot}}$ does not change with $a$ and remains at
315\,nm. This corresponds to the condition
$\text{Re}(\epsilon)\approx -1$ and is the intrinsic surface
plasmon resonance of a plane silver surface. As can be seen in the
plots, this resonance is strongly non-radiative, while the sphere
resonance is radiative in essence.}
 \label{FigMTotAlla}
\end{figure}

\subsection{Some simple examples}
\label{SecExamples}

The problem of an emitter close to a metal surface has been
studied extensively in the simple case of a plane surface with
various degrees of refinement
\cite{Chance78,Morawitz69,Kuhn70,Morawitz74,Fuchs81,Arnoldus88}.
Few studies have, however, focused on the regime of small
distances between emitter and metal (a few nanometers). In
principle, the non-local description of the dielectric function
fails at such small distances. However, because of the
difficulties of solving the non-local EM problem in complex
structures, it is useful to consider the local problem to obtain
qualitative and semi-quantitative results. Close to a plane,
$M_{\text{Tot}}$ can be very large, for example of the order of
$10^6$ at the Surface Plasmon (SP) resonance
($\text{Re}(\epsilon_r)\approx-1$) of noble metals. However, the
radiative enhancement is then (at best) of order$\sim$1, sometimes
much smaller. Most of the energy is therefore emitted into
non-radiative modes, and is eventually dissipated in the metal.
These modes have been shown to correspond to non-radiative Surface
Plasmon Waves (SPW). Although this identification requires a
non-local treatment, the study of the relative contribution of
radiative vs non-radiative decays is already contained in the
local approximation; this is precisely what is important for the
study of processes such as SE, SEF, and SERS. These studies on
plane metallic surfaces are the origin of the common belief that
fluorescence is always quenched close to metals. We will show here
that more complex metallic structures, in particular
nano-particles, can lead to enhancements with a large radiative
component.

\begin{figure}
\centering{
 \includegraphics[width=8cm]{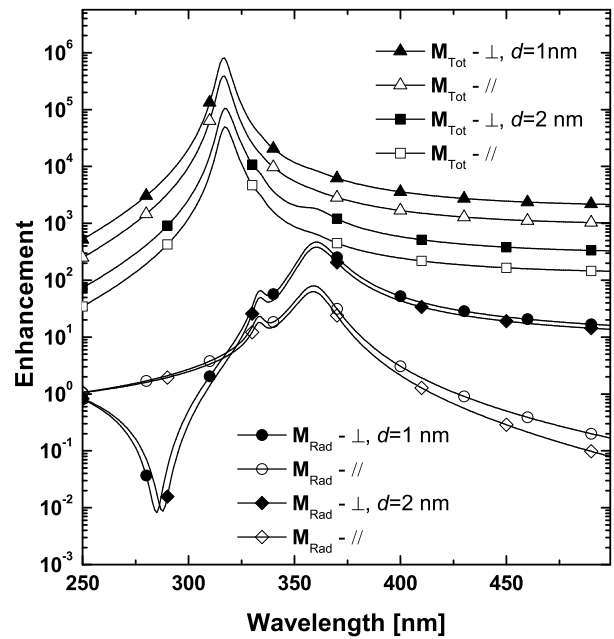}
} \caption{Total ($M_{\text{Tot}}$) and radiative
($M_{\text{Rad}}$) SE rate enhancements for a dipole close to a
silver sphere of radius $a=25$\,nm, as a function of wavelength.
The results are shown for dipoles either perpendicular or parallel
to the sphere surface, and at a distance $d=1$ or 2\,nm. The total
rate enhancements are strongly increased as the dipole gets closer
to the surface, while the radiative enhancements do not change
much.
 These plots were obtained using an extension of Mie theory, following Ref.
\cite{Chew87}. $M_{\text{Rad}}$ is obtained from the Poynting
vector approach, and $M_{\text{Tot}}$ from the self-reaction
approach. See the text for further details.}
 \label{FigMTotd1d2}
\end{figure}

To understand this, we show in Fig. \ref{FigMTotAlla} the
wavelength dependence of $M_{\text{Tot}}$ and $M_{\text{Rad}}$ for
a dipole perpendicular to the surface of a silver (Ag) sphere. The
results are shown for three sizes of spheres with radii $a=15$,
25, and 40\,nm, respectively; typical for colloidal Ag. Varying
the radius enables to understand the role of the main dipolar
surface plasmon resonance of the sphere, which red-shifts from
about 350 to 380\,nm as the size increases. $M_{\text{Rad}}$
clearly exhibits maxima at the SP resonances. On the opposite,
$M_{\text{Tot}}$ remains virtually unchanged for all sizes,
peaking at 315\,nm ($\text{Re}(\epsilon_r)\approx-1$),
corresponding to the intrinsic SP resonance of a plane Ag surface.
This can be understood simply: the sphere SP resonance is
radiative, and leads to a larger $M_{\text{Rad}}$ when the dipole
couples to it efficiently. This resonance is strongly {\it
geometry dependent}, and is red-shifted for larger objects, or for
coupled objects such as two closely-spaced spheres. However, the
intrinsic Ag SP resonance at 315\,nm is always strongly
non-radiative. It is related to the strong reflected field created
by the dipole image. This is {\it independent of geometry} because
the surface is approximately a plane when viewed from the dipole
at very short distances. The intensity of the reflected field is
however strongly dependent on the distance $d$ of the dipole from
the surface, and decreases as $d^{-3}$. This interpretation is
further confirmed in Fig. \ref{FigMTotd1d2} where the results for
two distances, $d=1$ and 2\,nm, are compared. The expected
dependence of $M_{\text{Tot}}$ with distance is clearly observed,
while $M_{\text{Rad}}$ remains virtually unchanged. This reflects
the fact that $M_{\text{Tot}}$ is dominated by the dipole image
reflected field (strongly distance dependent), while
$M_{\text{Rad}}$ depends primarily on how well the dipole emission
couples to the radiative SP resonance of the sphere. This coupling
is not strongly sensitive to distance, but should however be
sensitive to the dipole orientation. This is also illustrated in
Fig. \ref{FigMTotd1d2} where perpendicular and parallel dipoles
are compared.

Finally, we can also identify from Figs. \ref{FigMTotAlla} and
\ref{FigMTotd1d2} the situations where the radiative enhancement
is the largest. First, dipoles perpendicular to the surface
present larger enhancements than those parallel to it. Moreover,
as already pointed out, $M_{\text{Rad}}$ is maximum at the SP
resonance of the sphere, which is size-dependent. At this maximum,
$M_{\text{Rad}}$ is largest for the smallest sphere, with values
of $\approx 1100$ for $a=15$\,nm, down to $\approx 470$ for
$a=25$\,nm and $\approx 120$ for $a=40$\,nm. For a given radius,
$M_{\text{Rad}}$ varies little with $d$, down to $\approx 380$ for
$d=2$\,nm, or $\approx 210$ for $d=5$\,nm (for $a=25$\,nm).
However, because $M_{\text{Tot}}$ decreases strongly, the
radiative efficiency
$\eta_{\text{Rad}}=M_{\text{Rad}}/M_{\text{Tot}}$ increases
dramatically, from $\eta \approx 0.06$ for $d=1$\,nm to $\eta
\approx 0.5$ for $d=5$\,nm.

The previous discussion about metal spheres can be generalized to
more complex structures. However, to be quantitative, the solution
of the EM problem usually requires some approximations or the use
of numerical methods, which are outside the scope of this paper.
It is expected that suitable structures could lead to radiative
enhancements much larger than those for a sphere. For example, in
a recent study \cite{Blanco04}, numerical methods and the Poynting
vector approach were applied to closely spaced gold
nano-particles. Values of $M_{\text{Rad}}$ up to $2\times 10^5$
were calculated for a dipole placed at a junction between two
nano-particles. Here we are sacrificing examples with larger
enhancements for the convenience of having an analytic solution to
the problem through Mie theory.

\subsection{Using the Reciprocity Theorem}
\label{SecORT}

The two previous approaches require to solve Maxwell's equations
{\it for the dipolar singularity} in the presence of boundaries,
either numerically or analytically. This can present some
problems. Analytically, the dipolar singularity can significantly
complicate the problem (for example for a dipole near a sphere
\cite{Chew87}). Numerically, singularities are not always
straightforward to introduce. An additional inconvenience is that
the problem needs to be solved for each dipolar position and
orientation that one wants to study. We introduce here a third
approach, based on the optical reciprocity theorem, which can
circumvent some of these problems. The optical reciprocity theorem
(ORT) states \cite{Landau} that the field $\mathbf{E}$ created at
a given point M by a dipole $\mathbf{d}$ (at point O) is related
to the field $\mathbf{E}_2$ at O created by a dipole
$\mathbf{d}_2$ at M according to $\mathbf{d} \cdot
\mathbf{E}_2=\mathbf{d}_2 \cdot \mathbf{E}$. The validity is very
general, and in particular in the presence of absorbing
dielectrics or metals (assuming a local dielectric function). We
show here that this enables to derive the far-field properties of
the emitter in a given direction from the solution of two Plane
Wave Excitation (PWE) problems, without any source singularities.

We consider our dipole $\mathbf{d}$ at O in Fig.
\ref{FigSchemaSE}, and focus on its far-field emission at a
distance $R$ in direction $\mathbf{e}_r$ defined by angles
$(\theta,\phi)$. The radiation field is transverse (no radial
component) and can be decomposed into two components, $E_{\theta}$
and $E_{\phi}$, along unit vectors $\mathbf{e}_{\theta}$ and
$\mathbf{e}_{\phi}$. The Poynting vector is
$\mathbf{S}=(\epsilon_0 c / 2) |E|^2 \mathbf{e}_r$. To apply the
ORT, we consider first the problem of a dipole $\mathbf{d}_2= d_0
\mathbf{e}_{\theta}$ situated in M. The ORT yields $d_0
E_{\theta}=\mathbf{d} \cdot \mathbf{E}_2$, where $\mathbf{E}_2$ is
the field created by $\mathbf{d}_2$ at O. For sufficiently large
$R$, the field of this dipole in the region of interest can be
approximated by a plane wave propagating along $-\mathbf{e}_r$,
polarized along $\mathbf{e}_{\theta}$, and with amplitude $E_P=k^2
d_0 \exp(ikR) /(4\pi\epsilon_0 R)$.  We can now choose $d_0$ so
that $E_P=E_0$. The problem therefore reduces to solving Maxwell's
equation for excitation with a plane wave polarized along
$\mathbf{e}_{\theta}$, propagating along $-\mathbf{e}_r$, and of
amplitude $E_0$. We denote $\mathbf{E}^{\text{PW}-\theta}$ the
field at O for this plane wave problem and the ORT then yields for
the $\theta$ component of the radiation field of $\mathbf{d}$ at
M:
\begin{equation}
E_{\theta}=\frac{k^2 e^{ikR}} {4\pi\epsilon_0 R |E_0|} \mathbf{d}
\cdot \mathbf{E}^{\text{PW}-\theta},
\end{equation}
A similar expression is obtained for $E_{\phi}$, but note that it
requires the solution of a {\it different PWE problem} with
polarization along $\mathbf{e}_{\phi}$. The time-averaged power
radiated per unit solid angle is then
\begin{eqnarray}
\frac{dP_{\text{Rad}}}{d\Omega} (\theta,\phi)= R^2
\text{Re}(\mathbf{S} \cdot \mathbf{e}_r)=
 \frac{R^2 \epsilon_0 c}{2} {|E|^2}\\
 =\frac{\omega^4 d^2}{32 \pi^2 \epsilon_0 c^3 |E_0|^2} \left[
|\mathbf{e}_d \cdot \mathbf{E}^{\text{PW}-\theta}|^2+
|\mathbf{e}_d \cdot \mathbf{E}^{\text{PW}-\phi}|^2 \right].
\end{eqnarray}
It is easy to verify that the above expression is fully consistent
with that of an isolated dipole in free-space, yielding the same
expression as from a standard approach \cite{Jackson}. We can
therefore write the angle-dependent radiative enhancement as
\begin{equation}
M_{\text{Rad}} (\theta,\phi)
 =\frac{|\mathbf{e}_d \cdot \mathbf{E}^{\text{PW}-\theta}|^2}{|E_0|^2}+
\frac{|\mathbf{e}_d \cdot
\mathbf{E}^{\text{PW}-\phi}|^2}{|E_0|^2},
 \label{EqnMRadORT}
\end{equation}
 where $\mathbf{E}^{\text{PW}-\theta}$ and $\mathbf{E}^{\text{PW}-\phi}$
can be obtained from the solution of two different PWE problems.
The important result to highlight here is that {\it the radiative
property in a given direction of a dipole in a complex environment
can be obtained by modelling two PWE problems, without a dipolar
singularity}. Eq. (\ref{EqnMRadORT}) shows that the far-field
emission of a dipole in a given direction is in some way related
to the local field enhancement factor
$M_{\text{Loc}}=|E|^2/|E_0|^2$ for PWE from this direction. This
will be the basis for a the generalization of the $|E|^4$ SERS
enhancement in the next section.

\begin{figure}
\centering{
 \includegraphics[width=8cm]{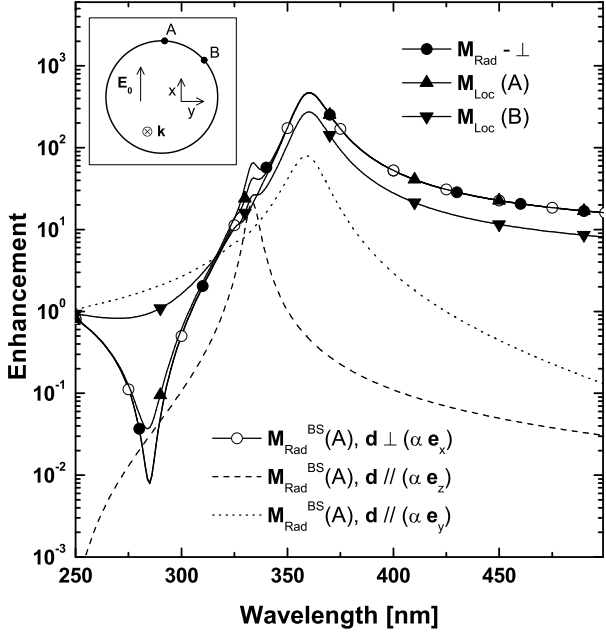}
} \caption{Example of calculations using the ORT approach for a
silver sphere of radius $a=25$\,nm and a dipole situated at
$d=1$\,nm from the surface at point A or B (see inset). The
radiative enhancement $M_{\text{Rad}}$ for a perpendicular dipole
and the local field enhancement $|E|^2/|E_0|^2$ at A and B are
shown for comparison. The angle dependent radiative enhancement is
calculated for the backscattering direction (along $-\mathbf{k}$)
for a dipole at A aligned along either $\mathbf{e}_x$
(perpendicular), or $\mathbf{e}_z$ or $\mathbf{e}_y$ (parallel).
This requires to calculate the field at A for two plane wave
problems, with polarization along $x$ and $y$. For the
perpendicular dipole, $M_{\text{Rad}}^{\text{BS}}$ is
indistinguishable from $M_{\text{Loc}}(\text{A})$ and
$M_{\text{Rad}}$. For other dipole orientations, it is clear that
$M_{\text{Rad}}^{\text{BS}}$ can be very different from
$M_{\text{Loc}}$.}
 \label{FigMRadRT}
\end{figure}

We illustrate the use of the ORT approach in Fig. \ref{FigMRadRT}.
The angle dependent radiative enhancement in direction
$-\mathbf{e}_z$ is calculated for dipoles located at point A (see
inset) with 3 different orientations. The results are compared to
calculations of the local field enhancement factor
$M_{\text{Loc}}=|E|^2/|E_0|^2$ for PWE along $\mathbf{e}_z$. Local
field and radiative enhancements are nearly identical for a
perpendicular dipole, but clearly differ for other orientations.
This highlights the difficulties of using purely a local field
enhancement approach to the problem of radiative enhancement. The
two approaches can be reconciled by the application of the ORT as
given in Eq. (\ref{EqnMRadORT}).

 We conclude this section with a few remarks:
\begin{itemize}
\item
 In principle, it is also possible within this approach to obtain
 the total radiative
enhancement $M_{\text{Rad}}$ by integrating $M_{\text{Rad}}
(\theta,\phi)$. However, it is often not practical and the
Poynting vector approach is usually better suited for this.
 \item
 This approach, contrary to the previous two, yields no information on
the total decay enhancement $M_{\text{Tot}}$ or decay to
non-radiative modes.
 \item
 This approach has hardly been used in the past \cite{Courtois96},
 but is actually well suited to many experimental situations where
 only radiative properties are of interest and where detection is performed
 in only one direction. We shall come back to it when
discussing SERS cross sections.
 \item
We would like to stress that the ORT as stated above does not
have, to our knowledge, any direct physical meaning. Its
expression, $\mathbf{d} \cdot \mathbf{E}_2=\mathbf{d}_2 \cdot
\mathbf{E}$, looks like a statement about interaction energies
between the two dipoles. However, one would need the complex
conjugates of the fields (or the dipoles) to be able to translate
this in a statement about the interaction energy of each dipole
with the field of the other. The demonstration of the ORT
\cite{Landau} does not involve, in fact, the interaction energy.
Moreover, the ORT considers the solution of two independent
problems with a single dipole and not the problem of the two
dipoles at the same time, thus avoiding mutual self-reactions. For
these reasons, the ORT has to be viewed as mathematical symmetry
relation embedded in Maxwell's equations, which relates the field
solution of two independent electromagnetic problems.
\end{itemize}

\subsection{Quantum yield of a fluorophore}
\label{SecQY}

Many applications make use of fluorophores, sometimes in complex
environments. There are two important characteristics of a
fluorophore: its absorption cross-section and its quantum yield,
both of which can be modified by the environment. Using the
previous arguments, we focus here on the modified quantum yield.
The absorption cross-section will be discussed later in the
context of Surface-Enhanced Fluorescence (SEF).

The fluorescence quantum yield is the proportion of the population
of excited electrons that decay radiatively to the ground state,
hence producing a detectable photon. It is related to the
radiative decay rate $\Gamma_{\text{Rad}}$, but other
(non-radiative) decay paths can exist in a molecule, even in
free-space. This intrinsic decay is characterized by a rate
$\Gamma_{\text{NR}}^{\text{int}}$. For an excited electron, there
is competition between these two mechanisms. In free-space, we
assume $\Gamma_{\text{Rad}}=\Gamma_0$ and the quantum yields is
then
\begin{equation}
Q_0=\frac{\Gamma_0}{\Gamma_0+\Gamma_{\text{NR}}^{\text{int}}}
\end{equation}

Close to a metal surface, there are two effects that will affect
the quantum yield: (i) the radiative decay rate
$\Gamma_{\text{Rad}}$ is modified by a factor $M_{\text{Rad}}$,
and (ii) there is an additional non-radiative path corresponding
to emission into dissipative modes in the metal, with rate
$\Gamma_{\text{NR}}=(M_{\text{Tot}}-M_{\text{Rad}}) \Gamma_0$. We
can rewrite the modified quantum yield as:
\begin{equation}
Q=\frac{\Gamma_{\text{Rad}}}{\Gamma_{\text{Rad}}+\Gamma_{\text{NR}}
+\Gamma_{\text{NR}}^{\text{int}}}
\end{equation}
Using the notations of the previous section and the expression for
$Q_0$, this can be expressed as
\begin{equation}
Q=\frac{M_{\text{Rad}}}{M_{\text{Tot}} +Q_0^{-1}-1}
 \label{EqnModQY}
\end{equation}

At short distances from the metal surface ($<10$ nm),
$M_{\text{Tot}}$ is typically much larger than 1. For a typical
fluorophore with a reasonable quantum yield, $M_{\text{Tot}}$ will
completely dominate the expression in the denominator. The
modified quantum yield becomes entirely governed by the
interaction with the metal, and {\it all fluorophores should
therefore exhibit the same quantum yield}. This aspect will be
discussed further in the context of SEF.

\section{SERS cross-sections}
\label{SecSERS}

\subsection{Scattering processes}

Scattering processes (elastic or inelastic) are, in many ways,
similar to SE. Excluding stimulation, the scattering involves the
`spontaneous' creation of a photon in a mode that was empty. One
expects a modification of the scattering cross-section in a
similar way as SE rates are modified by the environment. As
pointed out in the introduction, the difference between these two
processes lies in the initial state: $(i)$ excited atom with no
photon for SE and $(ii)$ ground state with one incident photon for
scattering. In the classical description of scattering, one can
formally distinguish between excitation and re-emission (even
though scattering is intrinsically one single process). Using the
concept of polarizability, $\alpha$, (for example Raman or
Rayleigh polarizabilities), the excitation part is described by
the creation of an induced dipole $\mathbf{d}=\alpha
\mathbf{E}_{\text{Loc}}$, where $\mathbf{E}_{\text{Loc}}$ is the
electric field at the scatterer position. The re-emission problem
of this dipole $\mathbf{d}$ can then be treated using the same
tools as that developed in Sec. \ref{SecSE}.

There is, however, a fundamental difference in the physical
interpretation of the enhancement factors.
$M_{\text{Tot}}=P_{\text{Tot}}/P_0$ characterizes the ability for
the environment to extract energy from the dipole, as compared to
free-space. For SE, this corresponds to a modification in the
total decay rate of the emitter. For scattering, because the
process (excitation+re-emission) is instantaneous, this
corresponds to a modification of the cross-section. We define here
the {\it total} scattering cross-section as
\begin{equation}
\sigma_{\text{Tot}}=P_{\text{Tot}}/S_0,
 \end{equation}
where
\begin{equation}
S_0=\epsilon_0 c |E_0|^2/2
\end{equation}
is the incident flux of the excitation wave (power per unit area),
and $P_{\text{Tot}}$ is the total power extracted from the induced
dipole, as in Sec. \ref{SecSE}. The standard definition of
scattering cross-section \cite{Jackson} corresponds to the {\it
radiative} scattering cross-section:
\begin{equation}
\sigma_{\text{Rad}}=\frac{P_{\text{Rad}}}{S_0},
\label{EqnSigRadGen}
\end{equation}
and similarly, the {\it differential radiative} scattering
cross-section:
\begin{equation}
\frac{d\sigma_{\text{Rad}}}{d\Omega}(\theta,\phi)=\frac{P_{\text{Rad}}(\theta,\phi)}{S_0}.
\end{equation}
We emphasize here the importance of distinguishing between
$\sigma_{\text{Rad}}$ and $\sigma_{\text{Tot}}$. The difference is
similar to the situation in SE between total and radiative decay
rates. $\sigma_{\text{Rad}}$ corresponds to the radiated scattered
field which can be measured experimentally in the far-field.
$\sigma_{\text{Tot}}$ includes $\sigma_{\text{Rad}}$, but in
addition accounts for non-radiative scattering events. These
events are not observable in the far-field, but they are indeed
experienced by the scatterer. For Raman scattering, it corresponds
therefore to the creation of a phonon, and to the emission of a
Stokes shifted photon, even if this photon is then absorbed by the
environment and not detectable in the far field. This will be
important later when discussing vibrational pumping in SERS. This
a similar situation as in modified SE, where an increase in the
total decay can be experienced by the emitter (and measured in
time-resolved measurements), despite the fact that only radiative
enhancement is measured in the far-field.

In Raman scattering, the scatterer is excited at a frequency
$\omega_L$, but the induced dipole oscillates at a different
frequency $\omega_R$. The difference in energy corresponds to the
creation (Stokes process) or destruction (Anti-Stokes process) of
a phonon (or vibration). The previous arguments can be easily
adapted to the presence of two different frequencies, one for
excitation ($\omega_L$), and one for re-emission ($\omega_R$).
This enables us to address specifically in the following the
SERS-EM enhancement in the context of SE.

\subsection{Local field enhancement factor}

In the presence of a dielectric or metallic environment, the field
$\mathbf{E}_{\text{Loc}}$ at the scatterer position is usually
different to the incident field $\mathbf{E_0}$. Under certain
conditions, in particular close to metallic surfaces, it is well
known that the local field intensity at the surface can be much
larger than that of the incident field, leading to a {\it local
field enhancement factor} $M_{\text{Loc}}(\omega_L)$. This factor
is present for all linear optical processes, such as absorption or
scattering. For Raman scattering, it is related to the excitation
of the Raman dipole and will depend on the exact Raman tensor of
the probe. We only consider here two simple and common cases. For
an isotropic Raman tensor, the induced Raman dipole is simply
$\mathbf{d}=\alpha \mathbf{E}$, and
\begin{equation}
M_{\text{Loc}}=|\mathbf{E}_{\text{Loc}}|^2/|E_0|^2, \label{mloc1}.
\end{equation}
For a uni-axial tensor, with a fixed axis along $\mathbf{e}_d$,
the induced dipole is then $\mathbf{d}=\alpha(\mathbf{e}_d \cdot
\mathbf{E})\mathbf{e}_d$ and
\begin{equation}
M_{\text{Loc}}=|\mathbf{e}_d \cdot
\mathbf{E}_{\text{Loc}}|^2/|\mathbf{e}_d \cdot \mathbf{E}_0|^2.
\label{mloc2}
\end{equation}

This enhancement has been studied for many geometries using
various approximations. Large values of $M_{\text{Loc}}$ are
associated with localized surface plasmon resonances at certain
wavelengths. They are also typically highly localized, i.e at the
junction of two closely spaced nano-particles. We showed in Sec.
\ref{SecORT} that these resonances are closely related to the
radiative SP resonances observed for $M_{\text{Rad}}$.
 $M_{\text{Loc}}$ can be
estimated to be $\sim 100-1000$ at the surface of an Ag colloid at
the plasmon resonance energy. Larger values ($10^5-10^6$) have
been predicted at the junction of two particles, or at sharp
corners. For interacting particles the collective resonance is
also red-shifted.

\subsection{SE approach to SERS cross sections}

SERS corresponds to Raman scattering in close proximity to
metallic surfaces, in particular, silver (Ag) and gold (Au). Under
the right conditions, a large enhancement is observed in the Raman
signal, with apparent cross-sections being as large as $10^{15}$
times larger than the normal cross-section. The local field
enhancement factor $M_{\text{Loc}}$ can be as large as
$10^5-10^6$, but such values are still too small to explain the
large reported SERS cross sections. The problem is that
$M_{\text{Loc}}$ describes well the enhancement associated with
the excitation of the molecule, but not the re-emission process.
The most common assumption in SERS is that $\sigma \propto
M_{\text{Loc}}(\omega_L)M_{\text{Loc}}(\omega_R)$, where
$M_{\text{Loc}}(\omega_R)$ is used to model the enhancement
associated with re-emission. This is the so-called
$|E|^4$-enhancement and enables to explain easily values up to
$10^{12}$. However, this approach has not been carefully justified
to the best of our knowledge. A few old studies use a proper
description of the dipolar emission in the vicinity of metal
surfaces, but analytical results are possible only for a single
(very small) spheres, where a relation of the type
$M_{\text{Loc}}(\omega_L)M_{\text{Loc}}(\omega_R)$ is obtained
\cite{Kerker80, Rojas93}. The first factor corresponds to
excitation while the second comes from the field emitted by the
dipole and scattered by the sphere. The optical reciprocity
theorem has been cited as a possible general justification of this
factor \cite{Xu04} without further details.

The treatment of SERS in the framework of modified SE modification
in fact shows that the SERS radiative cross-section enhancement
should be derived from Eq. (\ref{EqnSigRadGen}):
\begin{equation}
\frac{\sigma_{\text{Rad}}^{\text{SERS}}}{\sigma_0}=M_{\text{Loc}}(\omega_L)
M_{\text{Rad}}(\omega_R)
\label{EqnSERSRad}
\end{equation}
This seems in stark contrast with the standard $|E|^4$ approach.
The difference stems in the second (re-emission) term:
$M_{\text{Rad}}(\omega_R)$, instead of $M_{\text{Loc}}(\omega_R)$,
which can be calculated using the tools of Sec. \ref{SecSE}. These
two terms have {\it a priori} nothing in common. . The only
relation is through the orientation of the Raman dipole, which is
determined by the excitation and can affect the value of
$M_{\text{Rad}}(\omega_R)$. However, we will now show using the
optical reciprocity theorem that the two approaches can be
equivalent under certain conditions. This enables, accordingly, a
rigorous justification of the $|E|^4$ factor and defines precisely
its conditions of validity.

\subsection{Generalization of the $|E|^4$ factor}

In Sec. \ref{SecORT}, we used the ORT to give an alternative
expression for the radiative enhancement. The angle-dependent
radiative enhancement $M_{\text{Rad}}(\theta, \phi)$ was derived
from the solution of two plane wave excitation (PWE) problems. The
expression obtained in Eq. (\ref{EqnMRadORT}) resembles that of
$M_{\text{Loc}}$ in Eq. (\ref{mloc2}), and is the basis for the
justification and generalization of the $|E|^4$ factor. However,
because it applies to angle-dependent radiative enhancement, an
integration is required for the total radiative enhancement and
would lead to:
\begin{eqnarray}
\frac{\sigma_{\text{Rad}}^{\text{SERS}}}{\sigma_0}=M_{\text{Loc}}(\omega_L)
\times \nonumber\\
 \int \left[ \frac{|\mathbf{e}_d \cdot
\mathbf{E}^{\text{PW}-\theta}(\omega_R)|^2}{|E_0|^2}+
\frac{|\mathbf{e}_d \cdot
\mathbf{E}^{\text{PW}-\phi}(\omega_R)|^2}{|E_0|^2} \right]
d\Omega, \label{EqnSERSGen}
\end{eqnarray}
where $\mathbf{e}_d$ is the orientation of the Raman dipole
defined by the Raman tensor and the local field polarization. This
expression, although fairly general, has usually no interest in
practise because of its complexity. It would indeed require to
solve PWE problems for plane waves incident from all possible
directions $(\theta,\phi)$.

However, in most experimental situations, the SERS signal is
detected in the far field in only one given direction
$(\theta_d,\phi_d)$. The excitation and detection directions
depend on the specific scattering configuration and usually
correspond to a small solid angle defined by the numerical
aperture of the collecting optics. The physical quantity relevant
to the problem is therefore the differential radiative SERS
cross-section in the detection direction, $\sigma^d_{\text{Rad}}$.
Using Eq. (\ref{EqnMRadORT}), the SERS enhancement factor in this
direction is then given by
\begin{equation}
\frac{\sigma^d_{\text{Rad}}}{\sigma^d_0}=M_{\text{Loc}}(\omega_L)
M_{\text{Rad}} (\theta_d,\phi_d,\omega_R)
 \label{EqnSERSDiff}
\end{equation}

For simplicity, we will focus on the most common SERS setup,
namely the back-scattering (BS) configuration, where excitation
and detection are along the same direction, say (Oz). We also
assume that the incident wave is a plane wave polarized along
(Ox). As before, the use of the ORT requires to find the solution
of two plane wave problems, with polarization along (Ox) {\it and}
(Oy). We denote $\mathbf{E}^{\text{PX}}$ and
$\mathbf{E}^{\text{PY}}$ the electric fields at the scatterer
position for the solution of these two problems. We can then write
the BS radiative SERS enhancement factor as
\begin{equation}
\frac{\sigma^{\text{BS}}_{\text{Rad}}}{\sigma^{\text{BS}}_0}
=M_{\text{Loc}}(\omega_L) \left[ \frac{|\mathbf{e}_d \cdot
\mathbf{E}^{\text{PX}}(\omega_R)|^2}{|E_0|^2}+ \frac{|\mathbf{e}_d
\cdot \mathbf{E}^{\text{PY}}(\omega_R)|^2}{|E_0|^2} \right],
 \label{EqnSERSBS}
\end{equation}
which has some similarities with the conventional $|E|^4$ factor.
However, the standard $|E|^4$ factor, which would be here
proportional to $|E^{\text{PX}}(\omega_L)|^2
|E^{\text{PX}}(\omega_R)|^2$, in fact requires the solution of
only one plane wave problem, with polarization along (Ox). Since
PWE problems along (Ox) and (Oy) are two independent EM problems,
one concludes that there must be some information missing in the
$|E|^4$ approach, which is therefore at best an approximation.

The exact form of this expression can be further simplified by
considering specific Raman tensors for the probe. For an isotropic
Raman tensor, we have $\mathbf{d}=\alpha
\mathbf{E}^{\text{PX}}(\omega_L)$ leading to
\begin{eqnarray}
\frac{\sigma_{\text{Rad}}^{\text{BS}}}{\sigma^{\text{BS}}_0}=
\frac{|\mathbf{E}^{\text{PX}}(\omega_L)|^2
|\mathbf{E}^{\text{PX}}(\omega_R)|^2}{|E_0|^4} \nonumber\\
+ \frac{|\mathbf{E}^{\text{PX}}(\omega_L) \cdot
\mathbf{E}^{\text{PY}}(\omega_R)|^2}{|E_0|^4}
\end{eqnarray}
For the common case of a uniaxial Raman tensor, the induced Raman
dipole is $\mathbf{d}=\alpha (\mathbf{E}^{\text{PX}} \cdot
\mathbf{e}_d) \mathbf{e}_d$, where $\mathbf{e}_d$ is fixed by the
orientation of the molecule. The SERS enhancement is then
\begin{eqnarray}
\frac{\sigma^{\text{BS}}_{\text{Rad}}}{\sigma^{\text{BS}}_0}
=\frac{|\mathbf{e}_d \cdot \mathbf{E}^{\text{PX}}(\omega_L)|^2
|\mathbf{e}_d \cdot
\mathbf{E}^{\text{PX}}(\omega_R)|^2}{|\mathbf{e}_d \cdot \mathbf{E}_0|^2|E_0|^2} \nonumber \\
+\frac{|\mathbf{e}_d \cdot \mathbf{E}^{\text{PX}}(\omega_L)|^2
|\mathbf{e}_d \cdot
\mathbf{E}^{\text{PY}}(\omega_R)|^2}{|\mathbf{e}_d \cdot
\mathbf{E}_0|^2|E_0|^2}
\end{eqnarray}

From the derivation of Eq. (\ref{EqnMRadORT}), one actually sees
that the first term in the above expressions corresponds to
detection with a polarizer along (Ox), parallel to excitation,
while the second is that for a polarizer along (Oy), perpendicular
to excitation. The first term is equal to the conventional $|E|^4$
factor. We conclude that this factor can be exact, for example if
all the following conditions are met: (i) plane wave excitation,
(ii) back-scattering configuration, (iii) polarized detection
parallel to excitation, (iv) isotropic or uniaxial Raman tensor.
The reason for the first condition is actually quite natural. An
emitting dipole cannot distinguish the type of excitation being
used. Therefore, the re-emission part is always identical, say for
plane waves or Gaussian beams. However, the local field
enhancement can change from one type of excitation to another, and
the $|E|^4$ approach cannot be true for both. If condition (iii)
is not met, for example for non-polarized detection, the $|E|^4$
factor should be in most cases a good approximation, at least for
the order of magnitude of the enhancement. However, to study
depolarization effects, the standard $|E|^4$ factor is not
adequate and the full treatment given above is necessary.
Similarly, for other scattering configurations, for example at 90
degrees, the standard $|E|^4$ can lead to erroneous conclusions.
This emphasizes one important conclusion of this study often
overlooked in the past: accurate SERS enhancement calculations are
only possible if the Raman tensor of the probe and the scattering
configuration are clearly specified.

\subsection{SERS depolarization ratios and the $|E|^4$ factor}

One clear illustration of the failure of the $|E|^4$ factor is in
the calculation of SERS depolarization ratios. We here focus on
the simplest canonical example of a metallic silver sphere covered
by a continuous distribution of probe molecules on its surface.
Simple analytical expressions could be obtained in the
electrostatics approximation, but we give here exact results,
calculated using Mie theory \cite{Mie08, Bohren83}. Two situations
are considered: (i) molecules with an isotropic Raman tensor, and
(ii) molecules with a uni-axial Raman tensor. In the second case,
the axis needs to be specified, and we assume that for each
molecule it is fixed and perpendicular to surface of the sphere
(this corresponds to a fixed adsorption geometry). Choosing the
back-scattering geometry along direction (Oz), excitation is
polarized along (Ox) and detected with polarizer along (Ox)
(parallel) or (Oy) (perpendicular). The parallel and perpendicular
intensities are evaluated by integrating the intensities on the
sphere surface, to account for the contribution from all
molecules. According to the previous section, the depolarization
ratio $R=I_{\bot}/I_{\|}$ for the isotropic case is given by:
\begin{equation}
R^{\text{ORT}}= \frac {\int |\mathbf{E}^{\text{PX}}(\omega_L)
\cdot \mathbf{E}^{\text{PY}}(\omega_R)|^2} {\int
|\mathbf{E}^{\text{PX}}(\omega_L)|^2|\mathbf{E}^{\text{PX}}(\omega_R)|^2}.
\end{equation}
Using the conventional $|E|^4$ factor, combined with standard
techniques of depolarization scattering, one would obtain:
\begin{equation}
R^{\text{E4}}=\frac{\int
|\mathbf{E}^{\text{PX}}(\omega_L)|^2|\mathbf{e}_y \cdot
\mathbf{E}^{\text{PX}}(\omega_R)|^2} {\int
|\mathbf{E}^{\text{PX}}(\omega_L)|^2 |\mathbf{e}_x \cdot
\mathbf{E}^{\text{PX}}(\omega_R)|^2}.
\end{equation}

For the perpendicular uniaxial case, these relations are:
\begin{equation}
R^{\text{ORT}}= \frac {\int |\mathbf{E}^{\text{PX}}(\omega_L)
\cdot \mathbf{e}_r|^2 |\mathbf{E}^{\text{PY}}(\omega_R) \cdot
\mathbf{e}_r|^2} {\int |\mathbf{E}^{\text{PX}}(\omega_L) \cdot
\mathbf{e}_r|^2 |\mathbf{E}^{\text{PX}}(\omega_R) \cdot
\mathbf{e}_r|^2},
\end{equation}
and
\begin{equation} R^{\text{E4}}=\frac{\int
|\mathbf{E}^{\text{PX}}(\omega_L) \cdot \mathbf{e}_r|^2
|\mathbf{E}^{\text{PX}}(\omega_R) \cdot \mathbf{e}_r|^2
|\mathbf{e}_y \cdot \mathbf{e}_r|^2} {\int
|\mathbf{E}^{\text{PX}}(\omega_L) \cdot \mathbf{e}_r|^2
|\mathbf{E}^{\text{PX}}(\omega_R) \cdot \mathbf{e}_r|^2
|\mathbf{e}_x \cdot \mathbf{e}_r|^2}.
\end{equation}

\begin{figure}
\centering{
 \includegraphics[width=8cm]{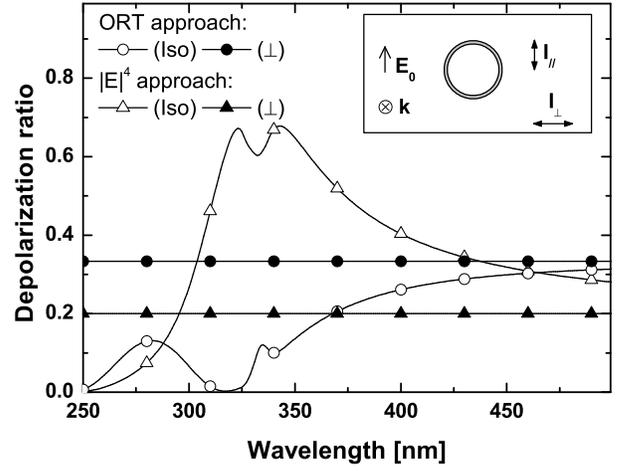}
} \caption{Comparison of the predictions from the standard $|E|^4$
approach, with that of modified SE using the ORT. The SERS
depolarization ratio $R=I_\bot /I_{\|}$ is shown as a function of
wavelength for a continuous distribution of probe molecules on the
surface of a silver sphere of radius $a=25$\,nm. Two Raman tensors
are considered: isotropic (Iso), or uniaxial with the main axis
perpendicular to the sphere surface ($\bot$). This clearly
illustrates the failure of the $|E|^4$ approximation under certain
conditions.}
 \label{FigDepol}
\end{figure}

For simplicity, we also make the common assumption of a negligible
Raman shift: $\omega_L \approx \omega_R$. Despite the apparent
similarities between the ORT and $|E|^4$ expressions, they are
actually very different. Note in particular the difference in the
numerators. In the first case, one needs to solve a second PWE
problem with polarization along (Oy) to obtain
$\mathbf{E}^{\text{PY}}$, while in the second, one PWE problem is
sufficient and the solution is simply projected onto
$\mathbf{e}_y$. This conceptual difference leads to drastically
different predictions, as shown in Fig. \ref{FigDepol}, where the
two models are plotted as a function of wavelength. In the
uniaxial case, the $|E|^4$ approach predicts a constant
(wavelength independent) ratio of $1/5$, whereas a ratio of $1/3$
is predicted from the ORT approach. For the isotropic tensor, the
behavior at the surface plasmon resonance of the sphere is simply
the opposite. The standard $|E|^4$ approach predicts a ratio
increasing to $\approx 0.7$, while the generalized approach
predicts a decrease close to 0. Such discrepancies are also likely
for depolarization ratios calculations in more complex geometries;
a subject which will be explored in further detail in a
forthcoming publication.

\subsection{Image dipole enhancement}

Before discussing these results any further, we would like to
mention an additional enhancement mechanism, which we have
neglected in the previous considerations, namely: the image dipole
enhancement. This does not apply to SE, but is present for many
other processes, in particular absorption and scattering. This
mechanism has been discussed in the past
\cite{King78,Efrima79,Weber80}, but has since then been
overlooked. It actually appears naturally when considering SERS in
the framework of modified spontaneous emission. For SE, the
self-reaction field modifies the ability of the dipole to radiate
energy. Although the `re-emission' enhancement in SERS follows the
physics of modified SE we have to consider an additional
ingredient: in SE the dipole is already in an `excited' state with
a fixed dipole amplitude, whereas for scattering, the dipole is
constantly driven by an external field. A constant source of
energy is then readily available from the laser. The self-reaction
field can therefore, in addition, oppose or amplify the dipole
amplitude (fixed by the dipole moment in SE). {\it This leads to
the concept of effective polarizability}. We assume in this
section for simplicity that the molecular polarizability and
self-reaction tensor $\mathbf{G}_r$ are isotropic. The dipole is
driven by both the external and reflected fields, which leads to a
modified polarizability of the form \cite{King78,Efrima79,Weber80}
\begin{equation}
\alpha_m=\alpha_0 \left( 1-\alpha_0 G_r \right) ^{-1},
\end{equation}
which in turn leads to an additional enhancement (or quenching)
factor
\begin{equation}
M_{\text{Im}}(\omega)=\frac{|\alpha_m|^2}{|\alpha_0|^2}=
\frac{1}{|1-\alpha_0(\omega) G_r(\omega)|^2}.\label{mim}
\end{equation}
For a dipole near a metallic plane, the self-reaction field in the
electrostatics approximation corresponds to the field created by
an image dipole and we have
\begin{equation}
G_r \approx \frac{1}{8\pi \epsilon_0 d^3}
\frac{\epsilon_r-1}{\epsilon_r+1},
\end{equation}
for a dipole perpendicular to the surface (half of this result for
a parallel dipole). $M_{\text{Im}}$ can therefore be called the
image dipole enhancement factor. This factor now involves both the
real and imaginary part of $G_r$.

The same effect applies to SERS through the effective Raman
polarizability. This has been studied in the past
\cite{King78,Efrima79,Weber80} and leads in the first
approximation to an additional enhancement
$M_{\text{Im}}(\omega_L)M_{\text{Im}}(\omega_R)$. The image dipole
enhancement factors had been proposed in the past to explain SERS
enhancements \cite{King78,Efrima79,Weber80} but were then thought
to be too small and have not been since then part of the main
discussions in the field. One reason is that it has so far been
studied mainly for plane metallic surfaces. In Sec.
\ref{SecExamples}, we have shown that for a sphere or more complex
structures, radiative enhancements can be large and form a
substantial part of $M_{\text{Tot}}$. From Eq. (\ref{EqnMTot}),
this means that $G_r$ must be strongly affected by the radiative
SP resonances. The common electrostatic approximation of the image
dipole then fails, and it is possible that image dipole
enhancement factors become non-negligible under these conditions.
This factor could then play a decisive role in explaining the role
of the polarizability of the probe in the total scattering cross
section and, in particular, the fact that dyes are the most
effective SERS probes in general.

\subsection{Comparison with Surface Enhanced Fluorescence}

\begin{figure}
\centering{
 \includegraphics[width=8cm]{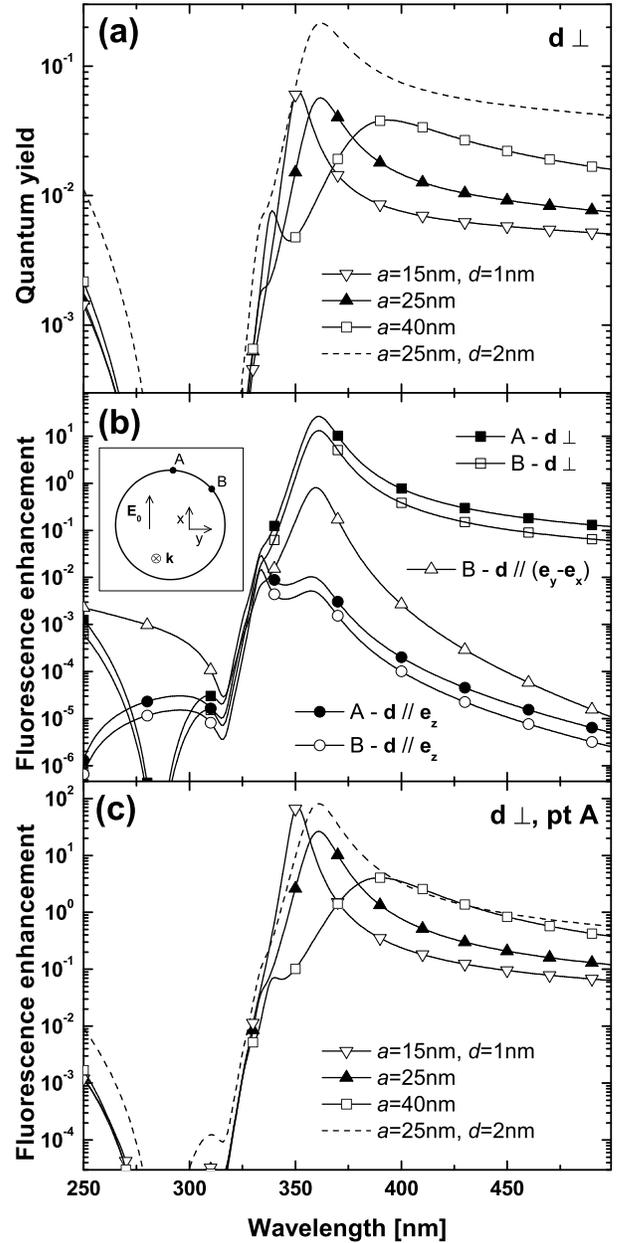}
} \caption{Surface Enhanced Fluorescence for a fluorophore
situated at a distance $d=1$ or 2\,nm from a silver sphere of
radii $a=15$, 25, and 40\,nm. The intrinsic quantum yield, $Q_0$
is assumed to be 1, but the results are insensitive to $Q_0$
unless it is extremely small. (a) Modified quantum yield (see Sec.
\ref{SecQY}). This is in principle not directly observable in SEF,
because local field enhancements also play an important role
during excitation of the fluorophore. (b) Fluorescence enhancement
for dipoles at two different position with various orientations.
Note that some configurations always result in quenching and not
enhancement. (c) Influence of $a$ and $d$ on the fluorencence
enhancement. Enhancements are observed at the surface plasmon
resonance of the sphere. Largest enhancements are observed for the
smallest sphere and for a dipole slightly away from the surface.
Note that for comparison with most experiments, it is necessary to
integrate the fluorescence intensity from all points on the
surface of the sphere.}
 \label{FigFluo}
\end{figure}

In many cases, the emission from fluorophores very close to metal
surfaces is strongly quenched by large non-radiative energy
transfer. However, several recent experiments have evidenced
fluorescence enhancements, for example close to Ag or Au surfaces
\cite{Lakowicz05,Song05}. There still seems to be some debate
about the origin of this enhancement and the exact conditions
under which it takes place. Many of the arguments and tools
developed so far can actually be adapted to the study of Surface
Enhanced Fluorescence (SEF), and shed light onto this problem.
This is also relevant to the problem of fluorescence quenching
under SERS conditions.

Fluorescence is very similar to scattering; it involves absorption
of a photon, followed by emission. The same enhancement mechanisms
as that derived for SERS are therefore expected. The absorption
should be subject to the local field enhancements, and the
emission is simply SE and should follow the same enhancements. The
image dipole enhancement should also be present for absorption,
but we neglect this aspect here for simplicity. {\it The crucial
difference is that SERS is instantaneous while fluorescence is a
multi-step process}. This means that both enhancements (local
field and SE type) contribute to the SERS cross-section. For
fluorescence, the local field enhancement does also lead to a
modification of the absorption cross-section, but the SE-type
enhancements only lead to modification of the decay rates
(radiative and non-radiative). To put it differently, once a
photon is absorbed and excites an electron in fluorescence,
enhancements cannot lead to more energy being extracted from this
electron, but only to energy being extracted faster.

To be more specific, we consider a fluorophore as in Sec.
\ref{SecQY}, with an absorption cross-section $k_0$, a radiative
rate $\Gamma_0$, and a quantum yield $Q_0$. We neglect here the
shift between excitation and fluorescence energies. Under a given
excitation, the photon flux is $n_0 \propto |E_0|^2$ and the power
radiated by fluorescence in free-space is (in number of photons
per unit time):
\begin{equation}
P_0^{\text{Fluo}}=k_0 n_0 Q_0
\end{equation}
We have assumed here that the power is small enough to avoid
saturation effects in the excited state. In this expression, $k_0
n_0$ is the number of photons absorbed per unit time. For each
excited electron, the probability of being emitted radiatively is
simply $Q_0$. If the fluorophore is now close to a metal surface,
we can use the same expression, with the modified quantum yield
$Q$ given by Eq. (\ref{EqnModQY}). The absorption rate is also
modified because the field intensity at the fluorophore position
is now $M_{\text{Loc}} |E_0|^2$. The fluorescence enhancement (or
quenching) is then:
\begin{equation}
\frac{P^{\text{SEF}}}{P_0^{\text{Fluo}}}=M_{\text{Loc}}
\frac{M_{\text{Rad}}}{M_{\text{Tot}} +Q_0^{-1}-1}
 \label{EqnSEF}
\end{equation}
Moreover, in many cases, $Q_0^{-1}-1$ can be neglected. This
expression clearly identifies the sources of enhancement or
quenching. There are in fact two competing mechanisms: (i) the
first term, the local field enhancement $M_{\text{Loc}}$, which in
most situations of interest is larger or much larger than 1, and
leads to enhanced fluorescence, and (ii) the second term which is
the radiative efficiency of the emitter in electromagnetic
interaction with the metal. This term is smaller (and sometimes
much smaller) than 1 and corresponds to quenching. Quenching and
enhancement of fluorescence are therefore two sides of the same
problem. The various reports on quenching and enhancement of
fluorescence are nothing but different situations where either the
first or second terms in Eq. (\ref{EqnSEF}) dominates with respect
to the other.

These aspects are illustrated in Fig. \ref{FigFluo}, for the case
of a silver sphere, where various cases of dipole orientation,
sphere radius, and distance from the surface are considered. These
plots show that the conditions for fluorescence enhancements
depend on many parameters, and are in general not trivial to
determine. There are however a few general observations:
\begin{itemize}
\item
 The region of fluorescence enhancement is relatively small
 compared to that of fluorescence quenching. It requires an
 efficient coupling to the radiative SP resonance of the sphere.
 This point had already been emphasized in Ref. \cite{Lakowicz05}.
 \item
 The largest enhancement can in principle be obtained for the
 smallest sphere radii (see Fig. \ref{FigFluo}(c)). This is the
 opposite conclusion to that reached in Ref. \cite{Lakowicz05},
 based on the idea that scattering is more dominant compared to
 absorption for the largest sphere. This re-emphasizes the fact
 that these effects are not simple and require a careful
 consideration of all enhancement and quenching factors.
 \item
 Increasing slightly the distance $d$ between emitter and surface in
 general leads to higher fluorescence enhancements. This is
 because the local field enhancement does not change much at short
 distances, while non-radiative decay decreases sharply with $d$.
 For dyes adsorbed on the surface, varying $d$ is not straightforward. However, for fluorescing quantum dots, $d$
 should depend on the size of the quantum dot, and a careful
 tuning could possibly lead to large fluorescence enhancements \cite{Song05}.
\end{itemize}

\section{Discussion}
\label{SecDisc}

In view of the previous description of SERS within the framework
of modified spontaneous emission, we can summarize the
electromagnetic enhancement of the radiative SERS cross section as
\begin{equation}
\frac{\sigma_{\text{Rad}}^{\text{SERS}}}{\sigma_0}=M_{\text{Im}}(\omega_L)M_{\text{Loc}}(\omega_L)
M_{\text{Rad}}(\omega_R) M_{\text{Im}}(\omega_R),
\label{EqnFullSERS}
\end{equation}
where $M_{\text{Im}}$ is given by Eq. (\ref{mim}),
$M_{\text{Loc}}$ is given by Eq. (\ref{mloc1}) (or Eq.
(\ref{mloc2}) depending on the Raman tensor), $M_{\text{Rad}}$ is
given by Eq. (\ref{EqnMRadORT}) (for detection along a specific
direction), and with all the expressions evaluated at at either
the laser $(\omega_L)$ or the Raman (Stokes or Anti-Stokes)
frequencies $(\omega_R)$.

The first two terms relate to the excitation part, while the last
two to re-emission. Moreover, $M_{\text{Rad}}$ is related to
$M_{\text{Loc}}$ for plane wave excitation using the ORT, in a way
defined in the previous sections. We conclude this study by
discussing briefly a few aspects of this expression that have
been, in our opinion, not fully accounted for in the past.

\subsection{Image dipole enhancement factors}

A detailed discussion of this factor requires a full paper in
itself. Our numerical simulations (not shown here) indicate that
this factor contributes to a $\sim 10^3$ enhancement at most. {\it
It is however important that this factor is the only one in} Eq.
(\ref{EqnFullSERS}) {\it that involves the linear polarizability
of the molecule}. It can therefore be the natural origin for a
number of characteristics in SERS. The fact that only dyes exhibit
the largest SERS cross section (having large optical
polarizabilities in the visible) comes as a natural consequence
here. We only mention briefly too that this factor can explain the
apparent resonance profile observed in many SERS experiments
\cite{LeRuCAP05,Maher05} and can be of additional importance in
Surface Enhanced Resonant Raman Scattering (SERRS).

\subsection{Fluorescence quenching and the SERS continuum}

We consider here the common case of a dye under SERS conditions.
It is well known that fluorescence is strongly quenched when large
SERS signals are observed. This should in principle be derived
naturally within our framework. However, we have seen that SERS
and SEF are in fact subject to the same local field enhancement
factor and one could therefore expect the two to be correlated. In
fact, combining Eqs. (\ref{EqnSERSRad}) and (\ref{EqnSEF}) we see
that the SERS and SEF enhancement factors are directly related
through $M_{\text{SERS}}=M_{\text{Tot}} M_{\text{SEF}}$. In a
typical SERS situation, $M_{\text{Tot}}$ can be very large, say
$10^5$, and SERS enhancements are therefore much larger than
fluorescence enhancements. This factor is still however much too
small to offset the intrinsic larger efficiencies of fluorescence
processes as compared to Raman, which typically differ by factors
of $\sim 10^{10}-10^{15}$. We would therefore expect fluorescence
to be enhanced less than Raman but still remain strong compared to
Raman under SERS conditions.

To understand this apparent paradox we need to re-examine the
situation of SEF under conditions of very high enhancement, where
the total decay rate $M_{\text{Tot}}\Gamma_0$ is many orders of
magnitude larger than usual. When decay rates become comparable or
smaller than the typical energy relaxation times, then the
electron dynamics in the relaxation pathways is strongly modified.
Very close to a noble metal surface $M_{\text{Tot}}$ can be $\sim
10^5$ or higher. For a typical SE lifetime of a few ns, the
modified lifetime would only be of the order of a few tens of fs
or less, i.e much shorter than typical energy relaxation times in
molecules. In this model, this would imply that absorbed photons
do not have time to relax and are emitted at the frequency they
were absorbed. Standard fluorescence would therefore be quenched
and appears as an increase in Rayleigh scattering. Indeed, such
enhancements have been reported under SERS conditions, but it is
difficult to test whether they are true Rayleigh scattering or
strongly modified fluorescence. Pushing this argument further, for
the shortest decay times, the uncertainty principle would also
imply that energy is not well defined in the intermediate levels
of the relaxation path with a typical spread of 1600~cm$^{-1}$ for
times of 10~fs. The whole process would qualitatively result in a
very broad emission from all possible intermediate virtual states.
We propose that this could be the origin of the strongly debated
SERS continuum, which is observed in many SERS experiments. A
proper treatment of such extreme conditions however requires a
full quantum mechanical study of absorption-relaxation-emission
similar to that of Ref. \cite{Xu04}.

\subsection{Non-radiative effects in SERS}

Finally, this approach also highlights the importance of
non-radiative effects in SERS. In fluorescence, non-radiative
effects manifest themselves as a reduced quantum yield, because of
the competition between radiative and non-radiative decays. In
SERS, because scattering is instantaneous, there is not such a
competition, and radiative and non-radiative emission can occur
independenly of each other. Similarly to SE, we can define a total
(radiative+non-radiative) SERS cross-section (we ignore here
$M_{\text{Im}}$).
\begin{equation}
\frac{\sigma_{\text{Tot}}^{\text{SERS}}}{\sigma_0}=M_{\text{Loc}}(\omega_L)
M_{\text{Tot}}(\omega_R)
\end{equation}
Under SERS conditions, this cross-section can be much larger than
the observable radiative SERS cross-section, as given in Eqs.
(\ref{EqnSERSRad}), (\ref{EqnSERSDiff}), or (\ref{EqnFullSERS}).
The difference corresponds to Raman processes resulting in
emission of a Raman photon in the non-radiative modes of the
metal. These Raman photons are obviously non-observable in the
far-field, but they do corresponds to a real Raman event, and
therefore to the creation or destruction of a vibration in the
molecule. This total SERS cross-section is therefore the one that
is relevant when studying phonon population dynamics, in
particular effects such as vibrational pumping
\cite{Kneipp96,LeRuFD132}. This could make the much debated
hypothesis of vibrational pumping more likely, since the
cross-section for phonon pumping could actually be much larger
than that inferred for the radiative SERS cross-sections in the
far-field.

\subsection{Conclusion}

In closing, we have presented a consistent formulation of
cross-section enhancements in SERS and fluorescence within the
framework of modified spontaneous emission. The local field
enhancements, and the competition between radiative and
non-radiative decays have been shown to have different
consequences for fluorescence and for scattering. This description
provides a relatively simple way of designing and modelling
metallic nano-structures for specific applications, either in SEF,
or in SERS. The model also highlights the shortcomings of the
usual $|E|^4$-proportionality in the SERS cross section, often
assumed in the literature, and provides a direct recipe for its
calculation based on classical electromagnetic theory,
self-reaction effects, and the optical reciprocity theorem. This
approach could explain some of the longstanding puzzles in SERS
like the presence of a SERS continuum, the increase in Rayleigh
scattering simultaneous with SERS, and the peculiar ability of
dyes to be very efficient probes. Full details of calculations
using this EM SERS cross section with applications to other more
complex and relevant geometries will be reported elsewhere.


\begin{thebibliography}{}
\bibitem{Purcell46}
 E. M. Purcell, Phys.\ Rev.\ {\bf 69},681
(1946).
\bibitem{Drexhage68}
 K. H. Drexhage, H. Kuhn, F. P. Sc\"{a}fer, Ber.\ Bunsenges.\
Phys.\ Chem.\ {\bf 72}, 329 (1968).
\bibitem{Goy83}
 P. Goy, J. Raymond, M. Gross, and S. Haroche, Phys.\ Rev.\ Lett.\
{\bf 50}, 1903 (1983).

\bibitem{Chance78}
 R. R. Chance, A. Prock, and R. Silbey, Adv. Chem. Phys. {\bf 37},
1 (1978).
\bibitem{Metiu84}
 H. Metiu, Progress. Surf. Sci. {\bf 17}, 153
(1984).
\bibitem{Ford84}
G. W. Ford and W. H. Weber, Phys. Rep. {\bf 113}, 195 (1984); and
references therein.

\bibitem{Dulkeith02}
E. Dulkeith, A. C. Morteani, T. Niedereichholz, T. A. Klar, J.
Feldmann, S. A. Levi, F. C. J. M. van Veggel, D. N. Reinhoudt, M.
M\"oller, and D. I. Gittins, Phys. Rev. Lett. {\bf 89}, 203002
(2002).

\bibitem{Lakowicz05}
J. R. Lakowicz, Analytical Biochem. {\bf 337}, 171 (2005).
\bibitem{Song05}
J.-H. Song, T. Atay, S. Shi, H. Urabe, and A. V. Nurmikko, Nano
Lett. {\bf 5}, 1557 (2005).

\bibitem{Moskovits85}
M. Moskovits, Rev. Mod. Phys. {\bf 57}, 783 (1985).

\bibitem{Kneipp97}
K. Kneipp, Y. Wang, H. Kneipp, L. T. Perelman, I. Itzkan, R. R.
Dasari, and M. S. Feld, Phys.\ Rev.\ Lett.\ {\bf 78}, 1667 (1997).
\bibitem{Nie97}
 S. Nie and S. R. Emory, Science {\bf 275}, 1102 (1997).

\bibitem{VoDinh99}
T. Vo-Dinh, D. L. Stokes, G. D. Griffin, M. Volkan, U. J. Kim, and
M. I. Simon, J. Raman Spectrosc. {\bf 30}, 785 (1999).
\bibitem{Kneipp02}
 K. Kneipp, H. Kneipp, I. Itzkan, R. R. Dasari, and M. S. Feld,
 J.\ Phys.: Cond.\ Matt.\ {\bf 14}, R597
(2002).
\bibitem{LeRuCAP05}
 E. C. Le Ru, M. Dalley, and P. G. Etchegoin , Curr. Appl. Phys. (in press).

\bibitem{Jiang03}
 J. Jiang, K. Bosnick, M. Maillard, and L. Brus, J. Phys. Chem. B {\bf
107}, 9964 (2003).
\bibitem{Xu04}
 H. Xu, X.-H. Wang, M. P. Persson, H. Q. Xu, M. K\"all, and P.
Johansson, Phys. Rev. Lett. {\bf 93}, 243002 (2004).

\bibitem{Blanco04}
L. A. Blanco and F. J. Garc\'{i}a de Abajo, Phys.\ Rev.\ B {\bf
69}, 205414 (2004).

\bibitem{Mie08}
 G. Mie, Ann.\ Phys.\ {\bf 25}, 377 (1908).

\bibitem{Landau} L. Landau, E. Lifchitz, and L. Pitaevskii, {\it Electromagnetics of Continuous
Media} (Pergamon, Oxford, 1984).

\bibitem{Bohren83}
 C. F. Bohren and D. R. Huffman, {\it Absorption
and Scattering of Light by Small Particles} (Wiley, New York,
1983).
\bibitem{Chew87}
H. Chew, J. Chem. Phys. {\bf 87}, 1355 (1987).

\bibitem{Barnett92}
S. M. Barnett, B. Huttner, and R. Loudon, Phys. Rev. Lett. {\bf
68}, 3698 (1992).

\bibitem{CDG1}
C. Cohen-Tannoudji, J. Dupont-Roc, and G. Grynberg, {\it Photons
and Atoms} (Wiley, New York, 1997),  p. 70.

\bibitem{Jackson}
J. D. Jackson, {\it Classical Electrodynamics} (Wiley, New York,
1998).


\bibitem{Morawitz69}
 H. Morawitz, Phys. Rev. {\bf 187}, 1792
(1969).
\bibitem{Kuhn70}
H. Kuhn, J. Chem. Phys. {\bf 53}, 101 (1970).
\bibitem{Morawitz74}
H. Morawitz and M. R. Philpott, Phys. Rev. B {bf 10}, 4863 (1974).
\bibitem{Fuchs81}
R. Fuchs and R. G. Barrera, Phys. Rev. B {\bf 24}, 2940 (1981).
\bibitem{Arnoldus88}
H. F. arnoldus and T. F. George, Phys. Rev. A {\bf 37}, 761
(1988).

\bibitem{Courtois96}
J.–Y. Courtois, J.–M. Courty, J. C. Mertz , Phys. Rev. A {\bf 53},
1862 (1996).



\bibitem{Kerker80}
M. Kerker, D.-S. Wang, and H. Chew, Appl. Optics {\bf 19}, 4159
(1980).
\bibitem{Rojas93}
R. Rojas V and F. Claro, J. Chem. Phys. {\bf 98}, 998 (1993).

\bibitem{King78}
F. W. King, R. P. Van Duyne, and G. C. Schatz, J. Chem. Phys. {\bf
69}, 4472 (1978).
\bibitem{Efrima79}
S. Efrima and H. Metiu, J. Chem. Phys. {\bf 70}, 1602 (1979).
\bibitem{Weber80}
W. H. Weber and G. W. Ford, Phys. Rev. Lett. {\bf 44}, 1774
(1980).
\bibitem{Maher05}
R. C. Maher, J. Hou, L. F. Cohen, E. C. Le Ru, J. M. Hadfield, J.
E. Harvey, P. G. Etchegoin, F. M. Liu, M. Green, R. J. C. Brown,
M. J. T. Milton, J. Chem. Phys., in press.

\bibitem{Kneipp96}
K. Kneipp, Y. Wang, H. Kneipp, I. Itzkan, R. R. Dasari, and M. S.
Feld, Phys. Rev. Lett. {\bf 76}, 2444 (1996).
\bibitem{LeRuFD132}
E. C. Le Ru and P. G. Etchegoin, Faraday Discuss. {\bf 132}, in
press.


\end{thebibliography}
\end{document}